# Evolution of generalized couple-stress continuum theories: a critical analysis

Ali R. Hadjesfandiari · Gary F. Dargush

Department of Mechanical and Aerospace Engineering
University at Buffalo, State University of New York, Buffalo, NY 14260, USA

ah@buffalo.edu, gdargush@buffalo.edu

December 29, 2014

**Abstract** In this paper, we examine different generalized couple-stress continuum mechanics theories, including couple stress, strain gradient and micropolar theories. First, we investigate the fundamental requirements in any consistent size-dependent couple stress continuum mechanics, for which satisfying basic rules of mathematics and mechanics are crucial to establish a consistent theory. As a result, we show that continuum couple stress theory must be based on the displacement field and its corresponding macrorotation field as degrees of freedom, while an extraneous artificial microrotation cannot be a true continuum mechanical concept. Furthermore, the idea of generalized force and independent generalized degrees of freedom show that the normal component of the surface moment traction vector must vanish. Then, with these requirements in mind, various existing couple stress theories are examined critically, and we find that certain deviatoric curvature tensors create indeterminacy in the spherical part of the couple stress tensor. We also examine micropolar and micromorphic theories from this same perspective.

# 1 Introduction

Classical continuum mechanics has provided a rational basis to analyze and understand the behavior of materials on human (or macro) scale for nearly two centuries, since the initial work of Poisson and Cauchy in the late 1820s. This theory contains no length scale parameter in its formulation and, hence, produces size independent solutions for all well-defined smooth problems. However, more recent experiments show that the mechanical behavior of materials in smaller scales is different from their behavior at the more familiar macro-scales. Therefore, further progress in micromechanics, nanomechanics and nanotechnology will require a consistent size-dependent continuum mechanics, which can account for the length scale effect due to the microstructure of materials. Furthermore, this size-dependent continuum mechanics can provide a more suitable connection to atomistic models and the fundamental base for developing size-dependent multi-physics formulations, such as those involving electro-mechanical coupling.

A review of the early literature reveals that classical or Cauchy continuum mechanics was based initially upon an atomistic representation of matter having only central forces among particles. As a result, the force-stresses $\sigma_{ij}$ describe the internal forces in the continuum model [1]. However, in a more realistic representation of matter the introduction of non-central forces in the underlying atomistic model is inevitable. This led Voigt [2] as a natural extension to consider also the effect of couple-stresses $\mu_{ij}$ in the corresponding continuum representation, although he did not develop a complete mathematical theory. In the first decade of the twentieth century, the Cosserat brothers [3] began to develop a mathematical model to analyze materials with couple-stresses. In the continuation of this development, the interaction in bodies is generally represented by true (polar) force-stress $\sigma_{ij}$ and pseudo (axial) couple-stress $\mu_{ij}$ tensors. The components of these force-stress and couple-stress tensors in the original form as identified by the Cosserats are shown in Fig. 1.

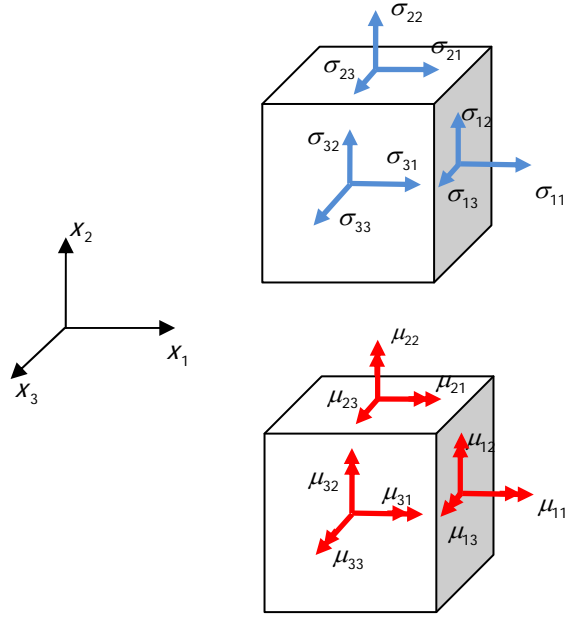

**Fig. 1** Components of force- and couple-stress tensors in the original Cosserat theory

As a result, the polar force-traction vector $t_i^{(n)}$ and axial moment-traction vector $m_i^{(n)}$ at a point on surface element $dS$ with unit normal vector $n_i$ are given by

$$t_i^{(n)} = \sigma_{ji} n_j, \tag{1}$$

$$m_i^{(n)} = \mu_{ji} n_j. \tag{2}$$

These vector tractions are shown in Fig. 2.

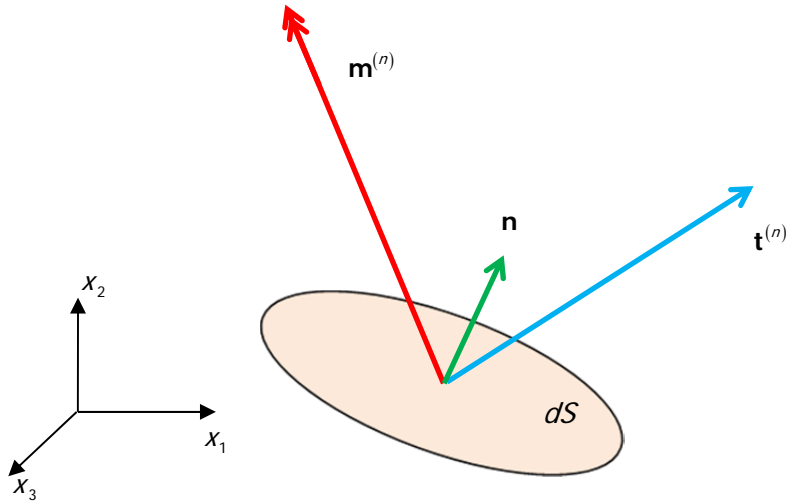

**Fig. 2** Force-traction $\mathbf{t}^{(n)}$ and moment-traction $\mathbf{m}^{(n)}$ system

As shown in Fig. 1, the force- and couple-stress tensors $\sigma_{ij}$ and $\mu_{ij}$ have eighteen components altogether. Since the tensors $\sigma_{ij}$ and $\mu_{ij}$ are general non-symmetric tensors, both can be decomposed into symmetric and skew-symmetric parts

$$\sigma_{ij} = \sigma_{(ij)} + \sigma_{[ij]}, \tag{3}$$

$$\mu_{ij} = \mu_{(ij)} + \mu_{[ij]}. \tag{4}$$

Here we have introduced parentheses surrounding a pair of indices to denote the symmetric part of a second order tensor, whereas square brackets are associated with the skew-symmetric part.

The linear and angular equilibrium equations in the original quasistatic Cosserat theory in differential form are given by

$$\sigma_{ji,j} + F_i = 0, \tag{5}$$

$$\mu_{ji,j} + \varepsilon_{ijk}\sigma_{jk} + C_i = 0, \tag{6}$$

where $F_i$ and $C_i$ are the body force and the body couple per unit volume of the body, respectively. Here $\varepsilon_{ijk}$ is the permutation tensor or Levi-Civita symbol. We notice that the angular equilibrium can be written as

$$\sigma_{[ji]} = \frac{1}{2}\varepsilon_{ijk}\mu_{lk,l} + \frac{1}{2}\varepsilon_{ijk}C_k, \tag{7}$$

which can be used to obtain the skew-symmetrical part of the force-stress tensor $\sigma_{[ji]}$. Therefore, the sole role of the angular equilibrium Eq. (6) is to produce the skew-symmetric part of the force-stress tensor. However, the appearance of body couple $C_i$ in Eq. (7) is very disturbing. After all, we expect the constitutive relation for $\sigma_{[ji]}$ be independent of the body couple $C_i$. This issue is even more serious in some dynamical models, in which the spin inertia distribution has also appeared.

As a result, the total force-stress tensor can be written as

$$\sigma_{ji} = \sigma_{(ji)} + \frac{1}{2}\varepsilon_{ijk}\mu_{lk,l} + \frac{1}{2}\varepsilon_{ijk}C_k. \tag{8}$$

By using this expression, the linear equation of equilibrium can be written as

$$[\sigma_{(ji)} + \frac{1}{2}\varepsilon_{ijk}\mu_{lk,l}]_{,j} + F_i + \frac{1}{2}\varepsilon_{ijk}C_{k,j} = 0. \tag{9}$$

This vector equilibrium equation involves fifteen components of stresses. Therefore, it requires twelve extra equations from constitutive relations; a task that does not look optimistic in this form.

It is obvious that by neglecting the effect of couple stresses and body couples

$$\mu_{ji} = 0 \quad , \quad C_i = 0, \tag{10a,b}$$

we obtain the classical or Cauchy continuum mechanics. In this classical theory, the angular equilibrium Eq. (6) or Eq. (7) show that the force-stress tensor is symmetric

$$\sigma_{[ji]} = 0, \quad \sigma_{ji} = \sigma_{ij} = \sigma_{(ji)}. \tag{11a,b}$$

This means that the tensor $\sigma_{ji}$ has six independent components and we have three linear equilibrium equations in Eq. (5). In the classical theory, the extra three equations are obtained by developing constitutive relations, which requires defining a measure of deformation in the material. In classical continuum mechanics, the deformation is specified by the displacement field $u_i$ and the consistent measure of deformation is the symmetric strain tensor $e_{ij}$, defined by

$$e_{ij} = \frac{1}{2}(u_{i,j} + u_{j,i}), \tag{12}$$

for infinitesimal deformations. However, this measure of deformation is not sufficient in the more realistic size-dependent continuum representation of matter, where there is no reason to neglect the effect of possible internal couple-stresses $\mu_{ij}$. We notice that in size-dependent couple stress continuum mechanics there are eighteen components of stresses, whereas we have six linear equilibrium equations in Eqs. (5) and (6). This makes the number of independent components twelve, which means the extra twelve equations must be obtained by developing constitutive relations. Therefore, size-dependent theories require introducing new degrees of freedom and new measures of deformations, in addition to the displacement vector $u_i$ and strain tensor $e_{ij}$. However, the development of so many different generalized couple-stress continuum theories in the last century shows that this has been a difficult task and, in particular, there has been confusion in defining kinematical degrees of freedom and measures of deformation. In hindsight, this indicates that something fundamental has been missed in the formulations. We should notice that the main difficulty in developing a consistent couple stress theory from the beginning has been the excessive number of components of force- and couple-stresses. As a result, we can suggest that there is a relation between the number of stress components and the description of deformation within a true continuum. This has been the main shortcoming until recent times.

Here we examine critically the evolution of the Cosserats' ideas toward the development of a consistent theory. This includes the original Cosserat theory [3], Mindlin and Tiersten [4], and Koiter [5] couple stress theories, Yang et al. [6] modified couple stress theory, strain gradient theories [7,8], as well as micropolar, microstretch and micromorphic theories [9-12]. Mainly based on the number of degrees of freedom and the additional measures of deformation, we demonstrate that these theories suffer from various inconsistencies. As a result, it has been impossible for any researcher to choose decisively which of these theories, if any, is self-consistent and worthy of further study. Metaphorically speaking, we might say that all these different theories have created a soup into which everybody adds some new ingredient based on his or her taste.

We organize the remainder of this paper as follows. In Sect. 2, we consider the original Cosserat theory and examine its shortcomings. Here we clarify the kinematics of a continuum by establishing the relation between displacement and rotation fields. Next, in Sect. 3, we consider the very important indeterminate couple stress theory developed by Mindlin and Tiersten and Koiter. By demonstrating the inconsistencies in this theory, we obtain the requirements for a consistent size-dependent couple stress theory. In Sect. 4, we examine the modified couple stress theory. Sect. 5 presents a review of general strain gradient theories. Then, in Sect. 6, we examine micropolar theory and its more generalized forms of microstretch and micromorphic theories, while Sect. 7 presents the recently developed consistent size-dependent couple stress theory. Sect. 8 contains a brief discussion and an overall comparison of the theories for linear elastic materials. Finally, Sect. 9 provides some general conclusions.

## 2 Original Cosserat theory

The Cosserat brothers [3] formulated several theories for structural elements, such as beams and shells, which represent one- and two-dimensional objects embedded in three-dimensional space. These theories involve displacements and independent rotational degrees-of-freedom in a natural way consistent with a continuum hypothesis. Based upon these successes, the Cosserats then extended this idea of independent rotational degrees-of-freedoms $\phi_i$ to the case of a full three-dimensional body in which each particle is outfitted with a triad of vectors, called directors [3]. Such formulations are today referred to as micropolar theories (Eringen [9], Nowacki [10]), which attempt to capture the effect of discontinuous microstructure by considering a continuous microrotation $\phi_i$ in addition to the translational degrees-of-freedom $u_i$. However, this extension is problematic, because it requires embedding a three-dimensional continuum, along with additional independent rotations $\phi_i$, into a three-dimensional space.

In a continuum representation of matter, it is assumed that matter is continuously distributed in space. As a result, the deformation of the body is represented by the continuous displacement field $u_i$ without considering the discontinuous microstructure of matter and motion of individual particles. This can be simply explained by considering an amount of gas in a closed container at thermodynamic equilibrium. In a continuum mechanical view, the velocity field of the gas is zero. However, we notice that the individual particles (molecules) have random motions (translations and rotations). The Maxwell–Boltzmann distribution in statistical mechanics describes particle speed probability density as a function of temperature. However, we notice that the average velocity or average momentum transfer is zero, which is consistent with a zero velocity field in continuum mechanics. When the gas flows, the continuum mechanical velocity field is actually the drift velocity at each point, which is the non-zero average velocity of the particles. We notice that there is the same analogy in fluids and solids, where the atoms have their vibrational and rotational motion around their equilibrium. However, these random motions are not considered in a continuum mechanical model. Here it should be mentioned that these random motions contribute to temperature and affect the material properties, such as heat capacity, viscosity and other material properties, which are used in developing constitutive relations in continuum mechanics. Therefore, the continuum mechanical velocity can be considered as a drift velocity added to the random motions. The temperature related random fluctuations (both translations and rotations) disappear in the kinematics of the continuum.

We notice that the geometrical points generally do not correspond to particles. As a result, the rotation field in a continuum is only defined based on relative motion of these geometrical points without considering the individual rotation of particles. Consequently, there is only one rotation field, derived from the displacement field $u_i$, defined by

$$\omega_i = \frac{1}{2}\varepsilon_{ijk}u_{k,j}. \tag{13}$$

This rotation was later called constrained rotation or macrorotation by proponents of micropolar theories, as opposed to microrotation $\phi_i$ in the original Cosserat theory. The concept of microrotation was introduced to account for the rotation of microelements or the director triads, which are different from the continuum mechanical rotation $\omega_i$. However, speaking of the rotation of microelements or individual particles in a continuum sense is meaningless, because we ignore the microstructure of matter after defining the displacement field. The motion of geometrical points is only represented by the continuous displacement field $u_i$ without considering the discontinuous microstructure of matter and motion of individual particles. Therefore, microrotation, which brings extraneous degrees of freedom, is not a proper continuum mechanical concept. How can the effect of the discontinuous microstructure of matter be represented mathematically by an artificial continuous microrotation? Thus, a consistent size-dependent couple stress continuum mechanics theory should involve only true continuum kinematical quantities, the displacement $u_i$, and its corresponding derived rotation $\omega_i$, without recourse to any additional artificial degrees of freedom. This means that the rigid body motion of infinitesimal elements of matter at each point of the continuum is described by six degrees of freedom, involving three translational $u_i$ and three rotational $\omega_i$ degrees of freedom.

Consequently, all micropolar formulations, which include an inconsistent independent continuous artificial microrotation $\phi_i$, suffer from this basic flaw.

The concept of independent rotational degrees of freedom $\phi_i$ may also originate from the discrete model of matter in molecular dynamics. In molecular dynamics, the lumped part of matter can be modeled as rigid bodies. As a result, the motion of each part can be described by motion of its center of mass and its rotation. In rigid body dynamics, this rotation is independent of the motion of the center of mass. Therefore, describing the translation and rotation of individual particles, such as atoms, molecules and grains, requires discrete point functions. On the other hand, the independent continuous artificial microrotation $\phi_i$ cannot represent these discrete point functions.

## 3 Indeterminate couple stress theory of Mindlin, Tiersten and Koiter (MTK theory)

Some researchers, such as Mindlin and Tiersten [4], and Koiter [5], speculated that in a consistent continuum theory, the deformation is completely specified by the continuous displacement field $u_i$. They considered that the kinematical quantities and measures of deformation are derived from this displacement field. Hence, in Mindlin-Tiersten-Koiter (MTK) theory, the rigid body motion of the infinitesimal element of matter at each point of the continuum is described by six degrees of freedom (i.e., three translational $u_i$ and three rotational $\omega_i$). As a result, energy considerations show that higher order measures of deformation must be related to the rotation field $\omega_i$.

We notice that for a material continuum occupying a volume $V$ bounded by a surface $S$, the principle of virtual work or weak formulation for equilibrium Eqs. (5) and Eq. (6) can be written as [13]

$$\int_V \sigma_{(ji)} \delta e_{ij} dV + \int_V \mu_{ji} \delta \omega_{i,j} dV = \int_S t_i^{(n)} \delta u_i dS + \int_S m_i^{(n)} \delta \omega_i dS + \int_V F_i \delta u_i dV + \int_V C_i \delta \omega_i dV . \qquad (14)$$

This relation shows that $\mu_{ji}$ and $\omega_{i,j}$ are energy conjugate tensors. Mindlin and Tiersten [4] and Koiter [5] considered the tensor $\omega_{j,i}$ as the infinitesimal curvature tensor, that is

$$\tilde{\kappa}_{ij} = \omega_{j,i}. \tag{15}$$

However, this curvature tensor creates some difficulties in the corresponding couple stress theory. First, we notice from Eqs. (13) and (15) that

$$\tilde{\kappa}_{ii} = \omega_{i,i} = 0, \tag{16}$$

which shows that the tensor $\tilde{\kappa}_{ij}$ is deviatoric, and similarly for the variations $\delta\omega_{i,j}$, and thus is specified by eight independent components. This character creates indeterminacy in the couple-stress tensor. This can be seen by decomposing the general tensor $\mu_{ij}$ into spherical $\mu_{ij}^{(S)}$ and deviatoric $\mu_{ij}^{(D)}$ parts in the following manner:

$$\mu_{ij} = \mu_{ij}^{(S)} + \mu_{ij}^{(D)}, \tag{17}$$

where

$$\mu_{ij}^{(S)} = \frac{1}{3}\mu_{kk}\delta_{ij}. \tag{18}$$

By denoting

$$Q = \frac{1}{3}\mu_{kk}, \tag{19}$$

we have

$$\mu_{ij}^{(S)} = \frac{1}{3}\mu_{kk}\delta_{ij} = Q\delta_{ij}. \tag{20}$$

Therefore, the couple-stress tensor can be written as

$$\mu_{ij} = Q\delta_{ij} + \mu_{ij}^{(D)}. \tag{21}$$

Accordingly, we notice that

$$\mu_{ji}\delta\omega_{i,j} = \mu_{ij}\delta\omega_{j,i} = \mu_{ij}^{(D)}\delta\tilde{\kappa}_{ij}. \tag{22}$$

This shows that it is actually the deviatoric part of the couple-stress tensor $\mu_{ij}^{(D)}$ that is energetically conjugate to the deviatoric curvature tensor $\tilde{\kappa}_{ij}$. As a result, we can only specify the deviatoric part $\mu_{ij}^{(D)}$ in this theory. In other words, the spherical part of the couple-stress tensor $\mu_{ij}^{(S)} = Q\delta_{ij}$ is indeterminate.

The indeterminacy of $Q$ then carries into the skew-symmetrical part of the force-stress tensor, such that

$$\begin{aligned}\sigma_{[ji]} &= \frac{1}{2}\varepsilon_{ijk}\mu_{lk,l} + \frac{1}{2}\varepsilon_{ijk}C_k \\ &= \frac{1}{2}\varepsilon_{ijk}Q_{,k} + \frac{1}{2}\varepsilon_{ijk}\mu_{lk,l}^{(D)} + \frac{1}{2}\varepsilon_{ijk}C_k.\end{aligned} \tag{23}$$

We should mention that the appearance of body couple $C_i$ in this equation is also a major issue. However, the indeterminacy of the couple stress tensor does not affect the force equilibrium Eq. (5), since

$$\begin{aligned}\sigma_{[ji],j} &= \frac{1}{2}\varepsilon_{ijk}Q_{,kj} + \frac{1}{2}\varepsilon_{ijk}\mu_{lk,lj}^{(D)} + \frac{1}{2}\varepsilon_{ijk}C_{k,j} \\ &= \frac{1}{2}\varepsilon_{ijk}\mu_{lk,lj}^{(D)} + \frac{1}{2}\varepsilon_{ijk}C_{k,j}.\end{aligned} \tag{24}$$

The other major difficulty in this development is the inconsistency of the boundary condition for the normal component of the moment traction. The right hand side of Eq. (14) shows that the boundary conditions on the surface of the body can be either vectors $u_i$ and $\omega_i$ as essential (geometrical) boundary conditions, or $t_i^{(n)}$ and $m_i^{(n)}$ as natural (mechanical) boundary conditions. This apparently makes a total number of six boundary values for either case. Consequently, there is no other possible type of boundary condition in size-dependent couple stress continuum mechanics. However, this is in contrast to the number of geometric boundary conditions that can be imposed [5]. In particular, if components of $u_i$ are specified on the boundary surface, then the normal component of the rotation $\omega_i$ corresponding to twisting

$$\omega_i^{(n)} = \omega^{(nn)} n_i = \omega_k n_k n_i. \tag{25}$$

where

$$\omega^{(nn)} = \omega_k n_k, \tag{26}$$

cannot be prescribed independently. Therefore, the normal component $\omega^{(nn)}$ is not an independent degree of freedom, no matter whether the displacement vector $u_i$ is specified or not. However, the tangential component of rotation $\omega_i$ corresponding to bending, that is,

$$\omega_i^{(ns)} = \omega_i - \omega_i^{(n)} = \omega_i - \omega_k n_k n_i, \tag{27}$$

represents two independent degrees of freedom in the global coordinate system, and may be specified in addition to $u_i$. As a result, the total number of geometric or essential boundary conditions that can be specified is five [5].

Next, we let $m_i^{(nn)}$ and $m_i^{(ns)}$ represent the normal and tangential components of the surface moment-traction vector $m_i^{(n)}$, respectively. The normal component

$$m_i^{(nn)} = m^{(nn)} n_i, \tag{28}$$

where

$$m^{(nn)} = m_k^{(n)} n_k = \mu_{ji} n_i n_j, \tag{29}$$

causes twisting, while

$$m_i^{(ns)} = m_i^{(n)} - m^{(nn)} n_i, \tag{30}$$

is responsible for bending. Therefore, the boundary moment surface virtual work in Eq. (14) can be written as

$$\int_S m_i^{(n)} \delta \omega_i dS = \int_S m_i^{(nn)} \delta \omega_i^{(n)} dS + \int_S m_i^{(ns)} \delta \omega_i^{(ns)} dS$$
$$= \int_S m^{(nn)} \delta \omega^{(nn)} dS + \int_S m_i^{(ns)} \delta \omega_i^{(ns)} dS. \tag{31}$$

As we know from theoretical mechanics, the generalized forces are associated only with independent generalized degrees of freedom, thus forming energetically dual or conjugate pairs. From the kinematic discussion above, $\omega^{(nn)}$ is not an independent generalized degree of freedom. Consequently, the corresponding generalized force must be zero and, for the normal component of the surface moment-traction vector $m_i^{(n)}$, we must enforce the condition

$$m^{(nn)} = m_k^{(n)} n_k = \mu_{ji} n_i n_j = 0 \quad \text{on } S. \tag{32}$$

Furthermore, the boundary moment surface virtual work in Eq. (31) becomes

$$\int_S m_i^{(n)} \delta\omega_i dS = \int_S m_i^{(ns)} \delta\omega_i dS = \int_S m_i^{(ns)} \delta\omega_i^{(ns)} dS. \tag{33}$$

This shows that a material in couple stress theory does not support independent distributions of normal surface moment (or twisting) traction $m^{(nn)}$, and the number of mechanical boundary conditions also is five. This result was first established by Koiter [5], although his couple stress theory does not satisfy this requirement.

To resolve this problem, Koiter [5] proposed, based on the Saint-Venant's principle, the possibility that a given $m^{(nn)}$ has to be replaced by an equivalent shear stress distribution and a line force system. He gave the detail analogous to the Kirchhoff bending theory of plates. However, there is a difference between couple stress theory and the Kirchhoff bending theory of plates, as we explain. It should be realized that Kirchhoff plate theory is once again a structural mechanics approximation to a continuum mechanics theory obtained by enforcing a constrained deformation. Consequently, results from this plate theory are not valid on and around the boundary surface, and near concentrated point and line loads. It is a fact that the plate theory usually gives better results in the internal bulk of the plate far enough from boundary and concentrated loads. However, couple stress theory is a continuum mechanics theory itself and should be valid everywhere, including near to and on the boundary, without any approximation. This means that a continuum theory should treat all parts of a material body with the same mathematical rigor and should not be considered as a structural mechanics formulation. Nevertheless, this fundamental difficulty with boundary conditions and its impact on the formulation was not appreciated at the time.

For further insight, we examine this inconsistent theory for small deformation elasticity. In an elastic material, there is an elastic energy density function $W$, where for arbitrary virtual deformations about the equilibrium position, we have

$$\delta W = \sigma_{ji} \delta e_{ij} + \mu_{ji} \delta \tilde{\kappa}_{ij}. \tag{34}$$

Therefore

$$W = W\left(e_{ij}, \tilde{\kappa}_{ij}\right). \tag{35}$$

Because of the deviatoric character of the curvature tensor, we notice that

$$\delta W = \sigma_{(ji)} \delta e_{ij} + \mu_{ji}^{(D)} \delta \tilde{\kappa}_{ij}, \tag{36}$$

which shows that

$$\sigma_{(ji)} = \frac{\partial W}{\partial e_{ij}}, \tag{37}$$

$$\mu_{ji}^{(D)} = \frac{\partial W}{\partial \tilde{\kappa}_{ji}}. \tag{38}$$

For general linear bi-anisotropic elastic material, the energy density function $W$ takes the form

$$W = \frac{1}{2} A_{ijkl} e_{ij} e_{kl} + \frac{1}{2} B_{ijkl} \tilde{\kappa}_{ij} \tilde{\kappa}_{kl} + C_{ijkl} e_{ij} \tilde{\kappa}_{kl}. \tag{39}$$

Here bi-anisotropic means that there is a cross-link relationship between $e_{ij}$ and $\tilde{\kappa}_{ij}$ through the tensor $C_{ijkl}$. However, when $C_{ijkl} = 0$, the material becomes anisotropic. The tensors $A_{ijkl}$, $B_{ijkl}$ and $C_{ijkl}$ contain the elastic constitutive coefficients and are such that the elastic energy is positive definite. As a result, tensors $A_{ijkl}$ and $B_{ijkl}$ are positive definite. We notice that the tensor $A_{ijkl}$ is actually equivalent to its corresponding tensor in Cauchy elasticity. Since the strain tensor $e_{ij}$ is symmetric and the curvature tensor $\tilde{\kappa}_{ij}$ is deviatoric, we have the symmetry relations

$$A_{ijkl} = A_{klij} = A_{jikl}, \tag{40}$$

$$B_{ijkl} = B_{klij}, \tag{41}$$

$$C_{ijkl} = C_{jikl}, \tag{42}$$

with constraints

$$B_{iikl} = 0, \quad B_{ijkk} = 0. \tag{43}$$

$$C_{ijkk} = 0. \tag{44}$$

These show that for the most general case, the number of distinct components for $A_{ijkl}$, $B_{ijkl}$ and $C_{ijkl}$ are 21, 36, and 48, respectively. Therefore, the most general linear elastic bi-anisotropic material is described by 105 independent constitutive coefficients. Since this theory requires very many material coefficients, it is less attractive for practical and experimental applications.

By using the energy density Eq. (39) in the general relations for stresses, Eqs. (37) and (38), we obtain the following constitutive relations

$$\sigma_{(ji)} = A_{ijkl} e_{kl} + C_{ijkl} \tilde{\kappa}_{kl}, \tag{45}$$

$$\mu_{ij}^{(D)} = B_{ijkl} \tilde{\kappa}_{kl} + C_{klij} e_{kl}. \tag{46}$$

As a result

$$\mu_{ij} = Q\delta_{ij} + B_{ijkl} \tilde{\kappa}_{kl} + C_{klij} e_{kl} \tag{47}$$

where again $Q$ is indeterminate. For linear isotropic elastic material, the symmetry relations require

$$A_{ijkl} = \lambda \delta_{ij} \delta_{kl} + \mu \delta_{ik} \delta_{jl} + \mu \delta_{il} \delta_{jk}, \tag{48}$$

$$B_{ijkl} = 4\eta \delta_{ik} \delta_{jl} + 4\eta' \delta_{il} \delta_{jk}, \tag{49}$$

$$C_{ijkl} = 0. \tag{50}$$

The moduli $\lambda$ and $\mu$ have the same meaning as the Lamé constants for an isotropic material in Cauchy elasticity. The material constants $\eta$ and $\eta'$ account for the couple-stresses in the isotropic material. As a result, the energy density takes the form

$$W = \frac{1}{2}\lambda e_{jj}e_{kk} + \mu e_{ij}e_{ij} + 2\eta \tilde{\kappa}_{ij}\tilde{\kappa}_{ij} + 2\eta'\tilde{\kappa}_{ij}\tilde{\kappa}_{ji}. \tag{51}$$

The following restrictions are necessary for positive definite energy density $W$

$$3\lambda + 2\mu > 0, \quad \mu > 0, \quad \eta > 0, \quad -1 < \frac{\eta'}{\eta} < 1. \tag{52}$$

The first two are identical to those from classical theory. As a result, we have the following constitutive relations for the symmetric part of the force-stress tensor and couple-stress tensor, respectively,

$$\sigma_{(ji)} = \lambda e_{kk}\delta_{ij} + 2\mu e_{ij}, \tag{53}$$

$$\begin{aligned}\mu_{ij} &= Q\delta_{ij} + 4\eta\tilde{\kappa}_{ij} + 4\eta'\tilde{\kappa}_{ji} \\ &= Q\delta_{ij} + 4\eta\omega_{j,i} + 4\eta'\omega_{i,j}.\end{aligned} \tag{54}$$

Then, by using Eq. (54) in Eq. (23), we obtain

$$\sigma_{[ji]} = \frac{1}{2}\varepsilon_{ijk}Q_{,k} + 2\eta\varepsilon_{ijk}\nabla^2\omega_k + \frac{1}{2}\varepsilon_{ijk}C_k, \tag{55}$$

for the skew-symmetric part of the force-stress tensor. Therefore, the total force-stress tensor becomes

$$\sigma_{ji} = \frac{1}{2}\varepsilon_{ijk}Q_{,k} + \lambda e_{kk}\delta_{ij} + 2\mu e_{ij} + 2\eta\varepsilon_{ijk}\nabla^2\omega_k + \frac{1}{2}\varepsilon_{ijk}C_k, \tag{56}$$

Notice the indeterminacy due to $Q$ and the presence of the body couple $C_k$.

Then, for the equilibrium equation in terms of the displacement, we obtain

$$\left(\lambda + \mu + \eta\nabla^2\right)u_{k,ki} + (\mu - \eta\nabla^2)\nabla^2 u_i + F_i + \frac{1}{2}\varepsilon_{ijk}C_{k,j} = 0. \tag{57}$$

The disappearance of the elastic constant $\eta'$ in the force-stress tensor $\sigma_{ji}$ in Eq. (56) and equilibrium Eq. (57) can be seen as the indication of inconsistency in this theory. Interestingly, the ratio

$$\frac{\eta}{\mu} = l^2, \tag{58}$$

specifies a characteristic material length $l$, which accounts for size-dependency. Thus, the final equations governing the isotropic linear solid in the small deformation couple stress elasticity theory under consideration can be written as

$$\sigma_{(ji)} = \lambda e_{kk}\delta_{ij} + 2\mu e_{ij}, \tag{59}$$

$$\mu_{ij} = Q\delta_{ij} + 4\mu l^2 \omega_{j,i} + 4\frac{\eta'}{\eta}\mu l^2 \omega_{i,j}, \tag{60}$$

$$\sigma_{ji} = \frac{1}{2}\varepsilon_{ijk}Q_{,k} + \lambda e_{kk}\delta_{ij} + 2\mu e_{ij} + 2\mu l^2 \varepsilon_{ijk}\nabla^2 \omega_k + \frac{1}{2}\varepsilon_{ijk}C_k, \tag{61}$$

$$\left[\lambda + \mu\left(1+l^2\right)\nabla^2\right]u_{k,ki} + \mu(1-l^2\nabla^2)\nabla^2 u_i + F_i + \frac{1}{2}\varepsilon_{ijk}C_{k,j} = 0. \tag{62}$$

Subsequently, Stokes [14] brought this formulation into fluid mechanics to model the size-dependency effect in fluids. It turns out that this is an interesting coincidence. George Gabriel Stokes generalized Navier equations for fluids, while much later Vijay Kumar Stokes brought MTK theory into fluid mechanics. However, as mentioned above, MTK couple-stress theory suffers from some serious inconsistencies and difficulties with the underlying formulations, which are summarized as follows:

1. The body-couple is present in the constitutive relations for the force-stress tensor in the MTK theory.

2. The spherical part of the couple-stress tensor is indeterminate, because the curvature tensor $\tilde{\kappa}_{ij} = \omega_{j,i}$ is deviatoric.

3. The boundary conditions are inconsistent, because the normal component of moment traction $m^{(nn)}$ appears in the formulation.

4. For linear bi-anisotropic elastic material, this theory requires 105 material constants, which makes the theory less attractive from both practical and experimental standpoints.

Interestingly, the indeterminacy of the spherical part of the couple stress tensor $Q\delta_{ij}$ in this inconsistent theory has been simply ignored without any reasonable justification in some work [15-22]. Eringen realized this indeterminacy as a major mathematical problem. As a result, he was the first to call this indeterminate couple stress theory [9]. In response to the appearance of this indeterminacy, Eringen and some other researchers returned to the original Cosserat theory and revived the idea of an independent artificial microrotation $\phi_i$ in developing many different micropolar theories. We will consider micropolar theory in Sect. 6.

As mentioned previously, for isotropic linear elastic material, the second elastic couple-stress constant $\eta'$ does not appear in the final governing equations. It turns out for the two-dimensional case, the stress boundary conditions are independent of $\eta'$; thus, the corresponding boundary value problem only depends on the first elastic couple-stress constant $\eta$. As a result, there have been many applications in the literature for two-dimensional isotropic elastic problems. However, for three-dimensional problems, this theory requires both couple stress material constants $\eta$ and $\eta'$. For example, in the torsion of a cylinder, this theory predicts appearance of couple-stresses [5], which depend on both constants $\eta$ and $\eta'$.

It should be noted that MTK theory is very influential in the history of couple-stress related theories. As will be seen, this theory has a direct impact in formulating the consistent couple stress theory. The most important advancement introduced in MTK theory was taking the continuous displacement vector $u_i$ as the fundamental variable to represent the deformation of the continuum domain. However, after developing the indeterminate couple stress theory [4], Mindlin himself was not entirely pleased with his formulation. This is obvious from the other formulations he developed, such as strain gradient theories [7,8] and micromorphic theory [11]. All of these developments suggest that perhaps Mindlin was not certain about the validity of any of his theories.

In a sharply critical passage in their definitive text on continuum mechanics, Truesdell and Noll [23] provided the following summary of the collective understanding, just after the shortcomings of MTK theory were realized: "The Cosserats' masterpiece stands as a tower in the field. Even the recent recreators of continuum mechanics, while they knew of it, did not know its contents in detail. Had they mastered it, not only would time and effort of rediscovery have been spared, but also a paragon of method would have lain in their hands." As we shall soon see, this judgment in favor of the Cosserat approach also was made in haste and the struggle to define a consistent couple stress theory remained for another half century.

In retrospect, the inconsistencies in MTK theory make one suspect that perhaps the number of independent stress components in a consistent couple stress size-dependent continuum mechanics theory should be less than eighteen. This is one reason why the theory with a symmetric couple-stress tensor was developed, which we consider in the following section.

## 4 Modified couple stress theory of Yang, Chong, Lam and Tong (YCLT theory)

Yang et al. [6] developed a model of couple stress, i.e., the modified couple stress theory, that considers an additional equilibrium equation for the moment of couple, in addition to the two equilibrium equations of the classical continuum. Application of this equilibrium equation, apparently leads to a symmetric couple-stress tensor, that is

$$\mu_{[ij]} = 0, \qquad \mu_{ij} = \mu_{ji}. \tag{63a,b}$$

As a result, the virtual work principle Eq. (14) shows that the symmetric part of $\omega_{i,j}$

$$\chi_{ij} = \frac{1}{2}\left(\omega_{i,j} + \omega_{j,i}\right), \tag{64}$$

is the corresponding curvature tensor in this theory. We notice that

$$\chi_{ii} = \omega_{i,i} = 0, \tag{65}$$

which shows that the tensor $\chi_{ij}$ is deviatoric, and thus is specified only by five independent components. As a consequence, all the inconsistencies in MTK theory, such as the indeterminacy in the couple-stress tensor and the appearance of $m^{(nn)}$ on the bounding surface, remain intact in this theory. We explore the details as follows.

First, we notice that the deviatoric character of $\chi_{ij}$ requires that

$$\mu_{ij}\delta\chi_{ij} = \mu_{ij}^{(D)}\delta\chi_{ij}, \tag{66}$$

which shows that the deviatoric part of the couple stress tensor $\mu_{ij}^{(D)}$ is energetically conjugate to the deviatoric curvature tensor $\chi_{ij}$. As a result, we can only specify the deviatoric part $\mu_{ij}^{(D)}$ of the couple-stress tensor in this theory and the spherical part of the couple-stress tensor $\mu_{ij}^{(S)} = Q\delta_{ij}$ is indeterminate. This indeterminacy appears in the skew-symmetrical part of the force-stress tensor, that is

$$\sigma_{[ji]} = \frac{1}{2}\varepsilon_{ijk}\mu_{lk,l} = \frac{1}{2}\varepsilon_{ijk}Q_{,k} + \frac{1}{2}\varepsilon_{ijk}\mu_{lk,l}^{(D)} + \frac{1}{2}\varepsilon_{ijk}C_k. \tag{67}$$

However, as in MTK theory, this indeterminacy does not affect the force equilibrium Eq. (5).

Since the couple-stress tensor $\mu_{ij}$ is symmetric here, this couple stress theory also does not satisfy the required boundary condition

$$m^{(nn)} = \mu_{ji}n_in_j = 0 \quad \text{on } S. \tag{68}$$

One might find recourse to Koiter's method to replace a given $m^{(nn)}$ by an equivalent shear stress and force system based on Saint-Venant's principle. However, again this is incompatible with the fact that the couple stress theory is a continuum theory, which should be valid everywhere, including the surface without any approximation.

For an elastic material in this theory, the elastic energy density function $W$ is defined, such that

$$W = W(e_{ij}, \chi_{ij}). \tag{69}$$

Therefore

$$\delta W = \sigma_{(ji)} \delta e_{i,j} + \mu_{ji}^{(D)} \delta \chi_{ij},\qquad(70)$$

which shows that

$$\sigma_{(ji)} = \frac{\partial W}{\partial e_{ij}},\qquad(71)$$

$$\mu_{ji}^{(D)} = \frac{\partial W}{\partial \chi_{ji}}.\qquad(72)$$

For linear bi-anisotropic elastic material, the energy density function $W$ takes the form

$$W = \frac{1}{2} A_{ijkl} e_{ij} e_{kl} + \frac{1}{2} B_{ijkl} \chi_{ij} \chi_{kl} + C_{ijkl} e_{ij} \chi_{kl}.\qquad(73)$$

The tensors $A_{ijkl}$, $B_{ijkl}$ and $C_{ijkl}$ contain the elastic constitutive coefficients and are such that the elastic energy is positive definite. As a result, tensors $A_{ijkl}$ and $B_{ijkl}$ are positive definite. Since the strain and curvature tensors are symmetric, we have the symmetry relations

$$A_{ijkl} = A_{klij} = A_{jikl},\qquad(74)$$

$$B_{ijkl} = B_{klij} = B_{jikl},\qquad(75)$$

$$C_{ijkl} = C_{jikl} = C_{ijlk},\qquad(76)$$

with constraints

$$B_{iikl} = 0,\qquad B_{ijkk} = 0,\qquad(77)$$

$$C_{ijkk} = 0.\qquad(78)$$

These show that for the most general case, the number of distinct components for $A_{ijkl}$, $B_{ijkl}$ and $C_{ijkl}$ are 21, 15, and 30, respectively. Therefore, the most general linear elastic bi-anisotropic material is described by 66 independent constitutive coefficients.

By using the energy density Eq. (73) in the general relations, Eqs. (71) and (72), we obtain the following constitutive relations

$$\sigma_{(ji)} = A_{ijkl} e_{kl} + C_{ijkl} \chi_{kl}, \tag{79}$$

$$\mu_{ij}^D = B_{ijkl} \chi_{kl} + C_{klij} e_{kl}. \tag{80}$$

Thus, we have

$$\mu_{ij} = Q \delta_{ij} + B_{ijkl} \chi_{kl} + C_{klij} e_{kl}, \tag{81}$$

For linear isotropic elastic material, the symmetry relations require

$$A_{ijkl} = \lambda \delta_{ij} \delta_{kl} + \mu \delta_{ik} \delta_{jl} + \mu \delta_{il} \delta_{jk}, \tag{82}$$

$$B_{ijkl} = 4\eta \delta_{ik} \delta_{jl} + 4\eta \delta_{il} \delta_{jk}, \tag{83}$$

$$C_{ijk} = 0. \tag{84}$$

The single material coefficient $\eta$ accounts for the couple-stresses in the isotropic material. As a result, the energy density takes the form

$$W = \frac{1}{2} \lambda e_{jj} e_{kk} + \mu e_{ij} e_{ij} + 4\eta \chi_{ij} \chi_{ij}. \tag{85}$$

The following restrictions are necessary for positive definite energy density $W$

$$3\lambda + 2\mu > 0, \quad \mu > 0, \quad \eta > 0. \tag{86}$$

Therefore, for linear isotropic elastic material, the constitutive relations reduce to

$$\sigma_{(ji)} = \lambda e_{kk} \delta_{ij} + 2\mu e_{ij}, \tag{87}$$

$$\begin{aligned} \mu_{ji} &= Q \delta_{ij} + 8\eta \chi_{ji} \\ &= Q \delta_{ij} + 4\eta \left( \omega_{i,j} + \omega_{j,i} \right), \end{aligned} \tag{88}$$

and the total force-stress tensor becomes

$$\sigma_{ji} = \frac{1}{2} \varepsilon_{ijk} Q_{,k} + \lambda e_{kk} \delta_{ij} + 2\mu e_{ij} + 2\eta \varepsilon_{ijk} \nabla^2 \omega_k + \frac{1}{2} \varepsilon_{ijk} C_k. \tag{89}$$

In this theory, for the equilibrium equation in terms of the displacement, we obtain

$$\left(\lambda+\mu+\eta\nabla^{2}\right)u_{k,ki} +(\mu-\eta\nabla^{2})\nabla^{2}u_{i} + F_{i} + \frac{1}{2}\varepsilon_{ijk}C_{k,j} = 0. \tag{90}$$

By using the characteristic material length $l$ defined by the ratio

$$\frac{\eta}{\mu} = l^{2}. \tag{91}$$

we can rewrite the constitutive and governing equations as

$$\sigma_{(ji)} = \lambda e_{kk}\delta_{ij} + 2\mu e_{ij}, \tag{92}$$

$$\mu_{ji} = Q\delta_{ij} + 8\mu l^{2}\chi_{ji}, \tag{93}$$

$$\sigma_{ji} = \frac{1}{2}\varepsilon_{ijk}Q_{,k} + \lambda e_{kk}\delta_{ij} + 2\mu e_{ij} + 2\mu l^{2}\varepsilon_{ijk}\nabla^{2}\omega_{k} + \frac{1}{2}\varepsilon_{ijk}C_{k}, \tag{94}$$

$$\left[\lambda+\mu\left(1+l^{2}\nabla^{2}\right)\right]u_{k,ki} + \mu(1-l^{2}\nabla^{2})\nabla^{2}u_{i} + F_{i} + \frac{1}{2}\varepsilon_{ijk}C_{k,j} = 0. \tag{95}$$

Interestingly, the force-stress tensor and the final governing equilibrium equations are similar to those in MTK theory for isotropic material. Therefore, we notice that the modified couple stress theory inherits all inconsistencies from indeterminate MTK theory. Nevertheless, the appearance of only one length scale parameter for isotropic material makes modified couple stress theory more desirable from an experimental and analytical view. As a result, this theory has been extensively used in many problems, such as bending, buckling and post-buckling, and vibration in recent years to investigate the mechanical behavior of the structures at small scale.

We notice that the results for two-dimensional isotropic problems are similar to those in indeterminate MTK theory. However, this theory is different for three-dimensional and anisotropic cases. Interestingly, for torsion of cylinder, this theory also predicts the appearance of couple stresses [6].

It should be emphasized that the modified couple stress theory cannot be taken as a special case of indeterminate MTK theory obtained by letting

$$\eta' = \eta. \tag{96}$$

This is obvious by noticing that this case is excluded by condition Eq. (52) for the indeterminate MTK couple stress theory. In addition, this similarity is only valid for isotropic material, and there is no simple analogy for general anisotropic and bi-anisotropic cases.

There have been some doubts about the validity of the modified couple stress theory from a different fundamental aspect. As mentioned, the symmetry character of the couple-stress tensor is the consequence of the peculiar equilibrium equation for the moment of couple, besides the two conventional equilibrium equations for force and couple. However, this requirement is an additional condition, which is not derived by any principle of classical mechanics, as mentioned by Lazopoulos [24]. Therefore, the modified couple stress theory not only inherits all inconsistences from indeterminate MTK theory, but also is based on an unsubstantiated additional artificial equilibrium of moment of couples in the set of fundamental equations. This new equilibrium equation has no physical explanation, and has been invented to make the couple-stress tensor symmetric. Let us examine this inconsistency in more detail.

As is known, a couple of forces is a free vector in conventional mechanics of rigid bodies. The motion of a rigid body is not affected by changing the position of a concentrated couple at point *A* to any arbitrary point *B* in the body. However, the same couple is not a free vector, when we analyze the distribution of the internal stresses and deformation of the body in a deformable continuum mechanics theory. In their development, Yang et al. [6] claim that the effect of a couple at a point *A* is equivalent to the effect of this couple at point *B* plus the moment of this couple about point *B*. However, this claim can be refuted very easily by realizing that the stresses and deformations of these two loadings cases are not the same. Interestingly, we expect that the concentrated couple creates singularity in stresses at point *A* for the first case, while it creates singularity at point *B* for the second case. These two different singularities indicate that the effects of these two couple systems are not the same. Furthermore, the governing equations in rigid body dynamics are based on the equations for the rate of change of linear momentum $\mathbf{P}$ and angular momentum $\mathbf{L}$ of the system of particles, i.e.

$$\sum \mathbf{F} = \frac{d\mathbf{P}}{dt}, \qquad (97)$$

$$\sum \mathbf{M} = \frac{d\mathbf{L}}{dt}, \qquad (98)$$

where $\sum \mathbf{F}$ and $\sum \mathbf{M}$ represent the sum of the external forces and the moment of external forces, respectively. Based on Newton's third law, the internal forces and their corresponding moments disappear in these relations. As a result, these fundamental equations establish that forces are sliding vectors and couples are free vectors, as long as the motion of rigid bodies are concerned. However, there is no additional analogous equation for the moment of angular momentum for system of particles in mechanics. This is because in this equation the effect of internal forces will not disappear. Therefore, Yang et al. [6] have violated fundamental laws of mechanics to make the couple stress tensor symmetric. Simplifying a theory might be acceptable, but only as long as it does not create fundamental inconsistencies.

We summarize the inconsistencies of the YCLT modified couple stress theory as follows:

1. The symmetric character of the couple-stress tensor $\mu_{ij}$ is based on an artificial fundamental law for equilibrium of moment of couples, which has no physical reality.

2. The spherical part of the couple-stress tensor is indeterminate, because the curvature tensor $\chi_{ij}$ is deviatoric.

3. The body-couple is present in the constitutive relations for the force-stress tensor.

4. The boundary conditions are inconsistent, because the normal component of moment traction $m^{(nn)}$ appears in the formulation.

## 5 General strain gradient theories

General strain gradient theories were also introduced in the 1960s by some researchers, including Mindlin [7], and Mindlin and Eshel [8]. In these theories, various forms of gradient of strain tensor have been taken as a fundamental measure of deformation. Interestingly, some forms of these theories have been also used in developing size-dependent multiphysics disciplines, such as flexoelectricity [25,26].

These theories describe the kinematics of the continuum by the displacement field $u_i$, as the fundamental variable. However, in these theories, the second gradient of deformations, such as $u_{i,jk}$ and $e_{ij,k}$ appear in the formulations explicitly. In some versions of these theories [7,27], the third gradient of displacement and higher order stresses are also introduced. However, we notice that although these theories utilize the continuous displacement vector $u_i$ as the fundamental variable to represent the deformation of the continuum, the second gradient of deformations $u_{i,jk}$ and $e_{ij,k}$ are not directly related to the rotation gradient $\omega_{i,j}$.

As mentioned previously, the left hand side of the virtual work principle Eq. (14)

$$\int_V \sigma_{(ji)} \delta e_{ij} dV + \int_V \mu_{ji} \delta \omega_{i,j} dV = \int_S t_i^{(n)} \delta u_i dS + \int_S m_i^{(n)} \delta \omega_i dS + \int_V F_i \delta u_i dV + \int_V C_i \delta \omega_i dV, \qquad (99)$$

shows that the general strain gradient $e_{ij,k}$ is not a fundamental measure of deformation in a consistent couple stress theory. This relation also shows that there is no room for the third and higher gradients of deformation in the formulation, because it also would require additional improper essential boundary conditions. Therefore, strain gradient theories are inconsistent continuum theories.

## 6 Micropolar, microstretch and micromorphic theories

Soon after realizing the inconsistency of indeterminate MTK couple stress theory, researchers also began to develop size-dependent theories that more closely resembled the Cosserats' original theory. The idea of microrotation $\phi_i$, a field independent of displacement field $u_i$, was again considered to be a fundamental kinematic quantity in an attempt to remedy the aforementioned issues with inconsistent indeterminate couple-stress theory. Eringen [9], Nowacki [10], Mindlin [11] and Eringen and Suhubi [12] were the first to revive various forms of the original Cosserat theory that now is more commonly referred to as micropolar, microstretch and micromorphic theories. In micropolar theories, to each point six degrees of freedom are attributed, which are represented by the displacement field $u_i$ for translational degrees-of-freedom and the microrotation field $\phi_i$ for the rotational degrees-of-freedom. However, as we explained before, the artificial microrotation $\phi_i$ as an independent variable is not compatible with the idea of a continuous medium and cannot describe the discontinuous microstructure of matter. Consequently, we expect that micropolar theories also create some inconsistencies in their formulation. For more clarification, we examine the micropolar theory of Eringen [9] deeply and find various new inconsistencies. Micropolar theories apparently seem to cure the indeterminacy problem, boundary condition problem and accommodate a place for body couple distribution [9]. However, we discover that these inconsistencies are transformed to new inconsistencies, as we now explain in more detail.

In micropolar theories, it is customary to assume that the moment-traction vector $m_i^{(n)}$ is energetically conjugate to the microrotation vector $\phi_i$ without any reasoning. As a result, the virtual work theorem in micropolar theories is apparently written as

$$\int_V \sigma_{ji} \left( \delta u_{i,j} - \varepsilon_{jik} \delta \phi_k \right) dV + \int_V \mu_{ji} \delta \phi_{i,j} dV = \int_S t_i^{(n)} \delta u_i dS + \int_S m_i^{(n)} \delta \phi_i dS + \int_V F_i \delta u_i dV + \int_V C_i \delta \phi_i dV . \quad (100)$$

However, careful examination shows that there is no reason why the macrorotation $\omega_i$ does not have any contribution to the work of the moment-traction $m_i^{(n)}$ on the surface in Eq. (100).

We notice that the left hand side of Eq. (100) shows that the tensors

$$\varepsilon_{ij} = u_{i,j} - \varepsilon_{jik} \phi_k , \quad (101)$$

$$k_{ij} = \phi_{i,j} , \quad (102)$$

are energy conjugate to the force-stress tensor $\sigma_{ji}$ and couple-stress tensor $\mu_{ji}$, respectively. As a result, these tensors are the measures of deformation in micropolar theory [9], where $\varepsilon_{ij}$ is the micropolar strain tensor and $k_{ij}$ is the micropolar curvature tensor. It is obvious that these tensors are generally non-symmetric and each is specified by nine independent components.

For an elastic material, the elastic energy density function $W$ becomes

$$W = W\left(\varepsilon_{ij}, k_{ij}\right). \tag{103}$$

Therefore

$$\delta W = \sigma_{ji}\delta\varepsilon_{ij} + \mu_{ji}\delta k_{ij}, \tag{104}$$

which apparently shows that

$$\sigma_{ji} = \frac{\partial W}{\partial \varepsilon_{ij}}, \tag{105}$$

$$\mu_{ji} = \frac{\partial W}{\partial k_{ij}}. \tag{106}$$

For linear bi-anisotropic elastic micropolar material, the energy density function $W$ takes the form

$$W = \frac{1}{2}A_{ijkl}\varepsilon_{ij}\varepsilon_{kl} + \frac{1}{2}B_{ijkl}k_{ij}k_{kl} + C_{ijkl}\varepsilon_{ij}k_{kl}. \tag{107}$$

The tensors $A_{ijkl}$, $B_{ijkl}$ and $C_{ijkl}$ contain the elastic constitutive coefficients and are such that the elastic energy is positive definite. As a result, tensors $A_{ijkl}$ and $B_{ijkl}$ are positive definite. Since the deformation tensors are general non-symmetric tensors, we have only the symmetry relations

$$A_{ijkl} = A_{klij}, \tag{108}$$

$$B_{ijkl} = B_{klij}. \tag{109}$$

These show that for the most general case, the number of distinct components for $A_{ijkl}$, $B_{ijkl}$ and $C_{ijkl}$ are 45, 45, and 81, respectively. Therefore, the most general linear elastic bi-anisotropic micropolar material is described by 171 independent constitutive coefficients. Thus, micropolar theories require more material parameters in constitutive relations than other theories we have examined thus far.

By using the energy density Eq. (107) in the general relations for stresses, Eqs. (105) and (106), we obtain the following constitutive relations

$$\sigma_{ji} = A_{ijkl}\varepsilon_{kl} + C_{ijkl}k_{kl},  \tag{110}$$

$$\mu_{ij} = B_{ijkl}k_{kl} + C_{klij}\varepsilon_{kl}. \tag{111}$$

For linear isotropic elastic micropolar material, the symmetry relations require

$$A_{ijkl} = \lambda\delta_{ij}\delta_{kl} + \mu\delta_{ik}\delta_{jl} + \mu'\delta_{il}\delta_{jk} \tag{112}$$

$$B_{ijkl} = \alpha\delta_{ij}\delta_{kl} + \beta\delta_{ik}\delta_{jl} + \gamma\delta_{il}\delta_{jk}, \tag{113}$$

$$C_{ijkl} = 0. \tag{114}$$

and the energy density becomes

$$W = \frac{1}{2}\lambda\varepsilon_{jj}\varepsilon_{kk} + \frac{1}{2}\mu\varepsilon_{ij}\varepsilon_{ij} + \frac{1}{2}\mu'\varepsilon_{ij}\varepsilon_{ji} + \frac{1}{2}\alpha k_{jj}k_{kk} + \frac{1}{2}\beta k_{ij}k_{ij} + \frac{1}{2}\gamma k_{ij}k_{ji}. \tag{115}$$

The following restrictions are necessary for positive definite energy density $W$

$$3\lambda + \mu + \mu' > 0, \quad \mu + \mu' > 0, \quad \mu' - \mu > 0, \tag{116}$$

$$3\alpha + \beta + \gamma > 0, \quad -\gamma < \beta < \gamma, \quad \gamma > 0. \tag{117}$$

As a result, we have the following constitutive relations for the force-stress and couple-stress tensors, respectively:

$$\sigma_{ji} = \lambda\varepsilon_{kk}\delta_{ij} + \mu\varepsilon_{ij} + \mu'\varepsilon_{ji}, \tag{118}$$

$$\mu_{ji} = \alpha k_{kk}\delta_{ij} + \beta k_{ij} + \gamma k_{ji}. \tag{119}$$

We notice that for linear isotropic micropolar elastic solids four additional constants are required, which presents a more difficult task from an experimental standpoint. Now we realize that the difficulties of indeterminate theory of MTK have been transformed to new troubles, such as the increased number of material properties.

By using these constitutive relations in the equilibrium Eqs. (5) and (6), we obtain the governing equations in terms of the displacement $u_i$ and microrotation $\phi_i$. However, the specification of boundary conditions is more problematic in this formulation. For example, as stated by Sadd [28], it is not clear how to specify the microrotation $\phi_i$ and/or moment-traction $m_i^{(n)}$ on the boundaries.

As mentioned, the indeterminacy of the couple-stress tensor mainly directed Mindlin and others to return to the Cosserats' original theory and develop new micropolar theories. What they did not realize is that the couple-stress tensor is still indeterminate, which we demonstrate in what follows.

Since $u_{i,j}$ is a true (polar) second order tensor, Eq. (101) shows that the micropolar strain tensor $\varepsilon_{ij}$ is a true (polar) second order tensor. This requires the microrotation $\phi_i$ to be a pseudo (axial) vector. As a result, the divergence of this vector, i.e., $\phi_{i,i}$, is a pseudo-scalar. This means $\phi_{i,i}$ changes sign under an inversion of the coordinate system. However, the appearance of a pseudo-scalar in the kinematics of a continuum is not possible, which means only scalar quantities exist. This requires that the divergence $\phi_{i,i}$ vanish, that is

$$\phi_{i,i} = k_{ii} = 0. \tag{120}$$

As a result, based on the Helmholtz decomposition theorem, the microrotation vector $\phi_i$ can be represented simply by the curl of a true (polar) vector field $\xi_i$, where

$$\phi_i = \frac{1}{2}\varepsilon_{ijk}\xi_{k,j}. \tag{121}$$

Here the factor ½ is arbitrarily added. Remarkably, this expresion for $\phi_i$ shows that the vector field $\xi_i$ can be taken as a displacement field, which can be conveniently called the microdisplacement field. This is the complement of our previous discussion that the introduction of micropolar rotation $\phi_i$ actually involves considering the continuity of matter for a second time, which creates inconsistency.

On the other hand, the compatibility Eq. (120) shows that the micropolar curvature tensor $k_{ij}$ is also deviatoric, and thus is specified only by eight independent components. As a result, the couple-stress tensor in micropolar theory is also indeterminate. The appearance of the microdisplacement field $\xi_i$ creates other speculations, such as its contribution as a virtual displacement in the expression of virtual work.

Interestingly, for general bi-anisotropic micropolar material, the condition Eq. (120) requires that

$$B_{iikl} = 0, \qquad B_{ijkk} = 0, \tag{122}$$

$$C_{ijkk} = 0. \tag{123}$$

These show that for the most general case, the number of distinct components for $A_{ijkl}$, $B_{ijkl}$ and $C_{ijkl}$ are 45, 36, and 72, respectively. Therefore, the most general linear elastic bi-anisotropic micropolar material, in what should be called indeterminate micropolar theory, is described by 153 independent constitutive coefficients. As a result, the constitutive relations become

$$\sigma_{ji} = A_{ijkl}\varepsilon_{kl} + C_{ijkl}k_{kl}, \tag{124}$$

$$\mu_{ji}^D = B_{ijkl}k_{kl} + C_{klij}\varepsilon_{kl}. \tag{125}$$

Therefore, micropolar theories have created new inconsistencies without resolving any of the former inconsistencies in MTK theory. We summarize inconsistencies of micropolar theories as follows:

1. The artificial microrotation $\phi_i$ is not a continuum mechanics concept. A continuous microrotation function cannot describe the rotation of individual point particles.

2. There is no reasoning for taking the degrees of freedom $u_i$ and $\phi_i$ as the energy conjugates of generalized tractions $t_i^{(n)}$ and $m_i^{(n)}$, respectively.

3. Since the microrotation is a pseudovector, the micropolar curvature tensor $k_{ij} = \phi_{i,j}$ is deviatoric. This makes the couple stress tensor indeterminate.

4. For linear bi-anisotropic elastic micropolar material, this theory requires 153 material constants. This makes the theory less attractive from both practical and experimental standpoints.

Interestingly, micropolar theories are considered as special cases of general microstretch and micromorphic theories [11, 12]. In this latter model, there are 12 additional degrees of freedom: three microdisplacement vector components and nine additional independent microdeformation tensor components. It is obvious these theories are also inconsistent.

Some micromorphic and micropolar theories also typically include in the dynamical case a spin inertia distribution, which cannot exist in a consistent continuum mechanical theory. Instead, we must notice that in consistent continuum mechanics the inertia of matter is only described by mass density. There is no such rotary inertia per unit volume in consistent theories of continuum mechanics. Therefore, the concept of microrotation $\phi_i$ and spin inertia distribution are the result of confusion caused by mixing discrete molecular dynamics with Cosserat continuum mechanics theory.

**7 Consistent size-dependent couple stress theory**

In this section, we present the consistent size-dependent couple stress theory, which ends the quest for a consistent Cosserat theory. This development involves only true continuum kinematical quantities without recourse to any additional artificial degrees of freedom. We notice that elements of developing this theory are based on MTK theory. Interestingly, we realize that the indeterminate MTK theory is an initial incomplete version of this consistent theory.

Based on our arguments in Sect. 2 and 3, we postulate that in a consistent continuum theory the deformation is only specified by the continuous displacement field $u_i$. Therefore, the rigid body motion of each infinitesimal element of matter at any point of the continuum is described by six degrees of freedom, involving three translational motion $u_i$, and three rotational $\omega_i$ components.

As mentioned, the principle of virtual work or weak formulation for equilibrium Eqs. (5) and (6) can be written as [13]

$$\int_V \sigma_{(ji)} \delta e_{ij} dV + \int_V \mu_{ji} \delta \omega_{i,j} dV = \int_S t_i^{(n)} \delta u_i dS + \int_S m_i^{(n)} \delta \omega_i dS + \int_V F_i \delta u_i dV + \int_V C_i \delta \omega_i dV. \quad (126)$$

This relation shows that a compatible curvature tensor must be related to $\omega_{i,j}$. However, in Sect. 3 and 4, we demonstrated that since this tensor and its symmetric part $\chi_{ij}$ are deviatoric, these cannot be considered as proper curvature tensors. Therefore, one might suggest that the skew-symmetric part of $\omega_{i,j}$

$$\kappa_{ij} = \frac{1}{2}(\omega_{i,j} - \omega_{j,i}), \quad (127)$$

is the consistent and proper curvature tensor, because its spherical part vanishes. Interestingly, in recent work [13], the present authors have demonstrated that this is actually the case by developing a consistent couple stress theory, which resolves the indeterminacy of the couple-stress tensor $\mu_{ji}$ and the issue with boundary conditions for the normal component $m^{(nn)}$ in MTK theory. Let us next examine this development in more detail.

In Sect. 3, by using arguments from Koiter [5], we demonstrated that the consistent couple stress theory requires that the normal surface moment-traction $m^{(nn)}$ on the boundary vanish, that is

$$m^{(nn)} = m_k^{(n)} n_k = \mu_{ji} n_i n_j = 0 \quad \text{on} \quad S. \tag{128}$$

However, as we discussed, this requirement was violated by former couple stress theories. It is important to notice that this fundamental difficulty with boundary condition and its impact on formulations was not understood before. Interestingly, this requirement has been fulfilled with remarkable consequences in Reference [13]. By using the concept of an arbitrary subdomain, we have shown that at any point in the body on any plane with normal direction $n_i$, $m^{(nn)}$ must vanish; that is

$$m^{(nn)} = \mu_{ji} n_i n_j = 0 \quad \text{in} \quad V. \tag{129}$$

Now since $n_i n_j$ is symmetric and arbitrary in Eq. (129), $\mu_{ji}$ must be skew-symmetric. Thus,

$$\mu_{(ij)} = 0, \quad \mu_{ij} = \mu_{[ij]}, \tag{130}$$

which means

$$\mu_{ji} = -\mu_{ij} \quad \text{in} \quad V. \tag{131}$$

This is the subtle character of the couple stress tensor in continuum mechanics, which has not been recognized by early investigators. We should notice that the skew-symmetric property of the couple-stress tensor is the result of its fundamental character, and has nothing to do with any constitutive relation. This property is valid for any solid, isotropic or anisotropic, elastic or inelastic, linear or non-linear. In this development, there are no additional assumptions beyond that of the continuum as a domain-based concept having no special characteristics associated with the actual bounding surface over any arbitrary internal surface.

Interestingly, the skew-symmetric character of the couple-stress tensor resolves the indeterminacy problem. We notice

$$\mu_{ii} = 0 \quad \text{in} \quad V, \tag{132}$$

which shows that $Q$ vanishes, and thus the couple-stress tensor $\mu_{ij}$ does not have any spherical part, that is

$$\mu_{ij}^{(S)} = Q \delta_{ij} = 0. \tag{133}$$

This means the couple-stress tensor is determinate in this new theory. The components of the force-stress $\sigma_{ij}$ and couple-stress $\mu_{ij}$ tensors in this consistent theory are shown in Fig. 3.

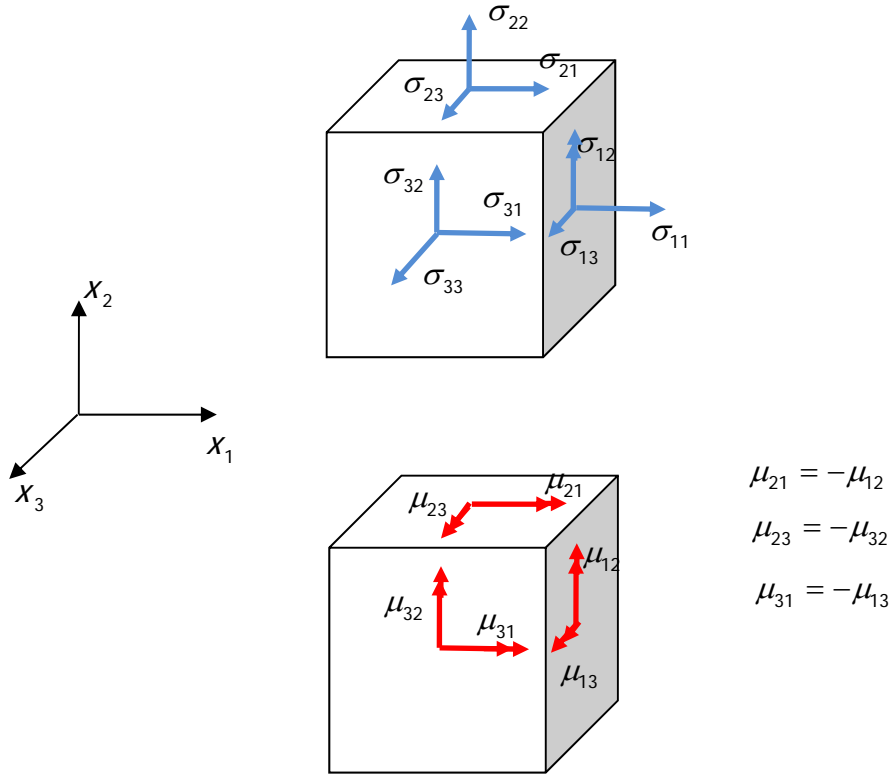

**Fig. 3** Components of force- and couple-stress tensors in the present consistent theory

Since $\mu_{ij}$ is skew-symmetric, the moment-traction $m_i^{(n)}$ given by Eq. (2) is tangent to the surface. As a result, the couple-stress tensor $\mu_{ij}$ creates only bending moment-traction on any arbitrary surface in matter. The force-traction $t_i^{(n)}$ and the consistent bending moment-traction $m_i^{(n)}$ acting on an arbitrary surface with unit normal vector $n_i$ are shown in Fig. 4.

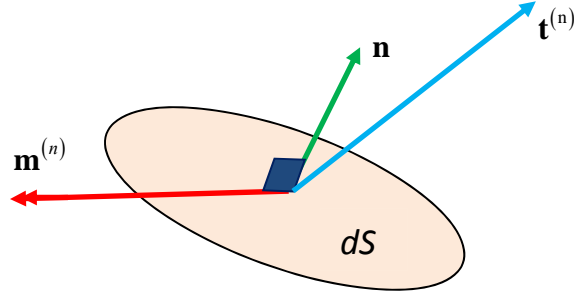

**Fig. 4** Force-traction $\mathbf{t}^{(n)}$ and the consistent bending moment-traction $\mathbf{m}^{(n)}$

The true couple-stress vector $\mu_i$ dual to the pseudo-tensor $\mu_{ij}$ is defined as

$$\mu_i = \frac{1}{2}\varepsilon_{ijk}\mu_{kj}. \tag{134}$$

Consequently, the surface moment-traction vector tangent to the surface $m_i^{(n)}$ reduces to

$$m_i^{(n)} = m_i^{(ns)} = \mu_{ji}n_j = \varepsilon_{ijk}n_j\mu_k. \tag{135}$$

Now we notice that the virtual work principle Eq. (126) shows that the skew-symmetric part of $\omega_{i,j}$

$$\kappa_{ij} = \frac{1}{2}\left(\omega_{i,j} - \omega_{j,i}\right), \tag{136}$$

is the consistent curvature tensor as suggested before. More inspection shows that pseudo-tensor $\kappa_{ij}$ is the mean curvature tensor, which represents the pure bending of material [13]. Interestingly, it turns out that the symmetric part of $\omega_{i,j}$

$$\chi_{ij} = \frac{1}{2}\left(\omega_{i,j} + \omega_{j,i}\right), \tag{137}$$

is the torsion pseudo-tensor representing the pure twist of material [13]. The skew-symmetric character of the couple-stress tensor shows that the symmetric torsion tensor does not contribute as a fundamental measure of deformation in a consistent couple stress theory.

Furthermore, the true mean curvature vector $\kappa_i$ dual to the pseudo-tensor $\kappa_{ij}$ is defined as

$$\kappa_i = \frac{1}{2}\varepsilon_{ijk}\kappa_{kj}. \tag{138}$$

By some manipulation, this can be written as

$$\kappa_i = \frac{1}{2}\omega_{ji,j} = \frac{1}{4}u_{j,ji} - \frac{1}{4}\nabla^2 u_i. \tag{139}$$

Accordingly, the mean curvature vector can be expressed in term of strain gradient as

$$\kappa_i = \frac{1}{2}e_{kk,i} - \frac{1}{2}e_{ik,k}. \tag{140}$$

Remarkably, the true continuum mechanical character of the pseudo-vector rotation $\omega_i$ resolves the issue with the appearance of body couple $C_i$ in Eq. (6). Reference [13] shows that the body couple virtual work in (12) can be transformed as

$$\int_V C_i \delta\omega_i dV = \int_V \frac{1}{2}\varepsilon_{ijk}C_{k,j}\delta u_i dV + \int_S \frac{1}{2}\varepsilon_{ijk}C_j n_k \delta u_i dS. \tag{141}$$

This indicates that the body couple $C_i$ transforms into an equivalent body force $\frac{1}{2}\varepsilon_{ijk}C_{k,j}$ in the volume and a force traction vector $\frac{1}{2}\varepsilon_{ijk}C_j n_k$ on the bounding surface. As a result, the body couple is not distinguishable from the body force. Therefore, in a proper couple stress theory, we must only consider body forces and the equilibrium equations become

$$\sigma_{ji,j} + F_i = 0, \tag{142}$$

$$\mu_{ji,j} + \varepsilon_{ijk}\sigma_{jk} = 0, \tag{143}$$

where

$$F_i + \frac{1}{2}\varepsilon_{ijk}C_{k,j} \to F_i \quad \text{in } V, \tag{144}$$

$$t_i^{(n)} + \frac{1}{2}\varepsilon_{ijk}C_j n_k \to t_i^{(n)} \quad \text{on } S. \tag{145}$$

Interestingly, in hindsight, we notice that the governing Eq. (57) in MTK theory shows these equivalent body forces. Finally, the virtual work theorem for consistent couple stress theory reduces to

$$\int_V \sigma_{(ji)}\delta e_{ij}\,dV + \int_V \mu_{ji}\delta\kappa_{ij}\,dV = \int_S t_i^{(n)}\delta u_i\,dS + \int_S m_i^{(n)}\delta\omega_i\,dS + \int_V F_i\delta u_i\,dV. \tag{146}$$

Since there is no indeterminacy in the couple-stress tensor in this development, for the skew-symmetric part of the force-stress tensor, we obtain

$$\sigma_{[ji]} = \frac{1}{2}\varepsilon_{ijk}\mu_{lk,l} = -\mu_{[i,j]}. \tag{147}$$

Thus, for the total force-stress tensor, we have

$$\sigma_{ji} = \sigma_{(ji)} + \frac{1}{2}\varepsilon_{ijk}\mu_{lk,l} = \sigma_{(ji)} - \mu_{[i,j]}. \tag{148}$$

Therefore, there are nine independent stress components in the consistent couple stress theory of size-dependent continuum mechanics. This includes six components of $\sigma_{(ji)}$ and three components of $\mu_i$. As a result, the linear equation of equilibrium reduces to

$$[\sigma_{(ji)} + \mu_{[j,i]}]_{,j} + F_i = 0, \tag{149}$$

which shows that there are are only three independent equilibrium equations. Therefore, we must obtain the necessary extra six remaining equations from constitutive relations.

For an elastic material, the elastic energy density function $W$ becomes

$$W = W(e_{ij}, \kappa_{ij}). \tag{150}$$

Therefore

$$\delta W = \sigma_{(ji)}\delta e_{i,j} + \mu_{ji}\delta\kappa_{ij}, \tag{151}$$

which shows that

$$\sigma_{(ji)} = \frac{\partial W}{\partial e_{ij}}, \tag{152}$$

$$\mu_{ji} = \frac{\partial W}{\partial \kappa_{ji}}. \tag{153}$$

For linear bi-anisotropic elastic material, the energy density function $W$ takes the form

$$W = \frac{1}{2} A_{ijkl} e_{ij} e_{kl} + \frac{1}{2} B_{ijkl} \kappa_{ij} \kappa_{kl} + C_{ijkl} e_{ij} \kappa_{kl}. \tag{154}$$

The tensors $A_{ijkl}$, $B_{ijkl}$ and $C_{ijkl}$ contain the elastic constitutive coefficients and are such that the elastic energy is positive definite. As a result, tensors $A_{ijkl}$ and $B_{ijkl}$ are positive definite. The tensor $A_{ijkl}$ is actually equivalent to its corresponding tensor in Cauchy elasticity. Since the strain tensor is symmetric and the mean curvature tensor is skew-symmetric, we have the symmetry relations

$$A_{ijkl} = A_{klij} = A_{jikl}, \tag{155}$$

$$B_{ijkl} = B_{klij} = -B_{jikl}, \tag{156}$$

$$C_{ijkl} = C_{jikl} = -C_{ijlk}, \tag{157}$$

These show that for the most general case, the number of distinct components for $A_{ijkl}$, $B_{ijkl}$ and $C_{ijkl}$ are 21, 6, and 18, respectively. Therefore, the most general linear elastic bi-anisotropic material is described by 45 independent constitutive coefficients.

By using the energy density Eq. (154) in the general relations, Eqs. (152) and (153), we obtain the following constitutive relations

$$\sigma_{(ji)} = A_{ijkl} e_{kl} + C_{ijkl} \kappa_{kl}, \tag{158}$$

$$\mu_{ij} = B_{ijkl} \kappa_{kl} + C_{klij} e_{kl}. \tag{159}$$

We notice that the skew-symmetric character of the couple-stress tensor $\mu_{ij}$ has a dramatic effect in reducing the number of independent coefficients $B_{ijkl}$ and $C_{ijkl}$ from 36 and 48 in the indeterminate MTK couple stress theory to 6 and 18, respectively.  This character has even more impact in advanced size-dependent modeling of continua in different branches of multi-physics disciplines.

For linear isotropic elastic material, the symmetry relations require

$$A_{ijkl} = \lambda \delta_{ij}\delta_{kl} + \mu \delta_{ik}\delta_{jl} + \mu \delta_{il}\delta_{jk}, \tag{160}$$

$$B_{ijkl} = 4\eta \delta_{ik}\delta_{jl} - 4\eta \delta_{il}\delta_{jk}, \tag{161}$$

$$C_{ijkl} = 0. \tag{162}$$

As a result, the energy density becomes

$$W = \frac{1}{2}\lambda e_{jj}e_{kk} + \mu e_{ij}e_{ij} + 4\eta \kappa_{ij}\kappa_{ij}. \tag{163}$$

The following restrictions are necessary for positive definite energy density $W$

$$3\lambda + 2\mu > 0, \quad \mu > 0, \quad \eta > 0. \tag{164}$$

The ratio

$$\frac{\eta}{\mu} = l^2 \tag{165}$$

specifies a characteristic material length $l$, which accounts for size-dependency in the small deformation couple stress elasticity theory under consideration here.

As a result, we have the following constitutive relations for the symmetric part of the force-stress and couple-stress tensors, respectively,

$$\sigma_{(ji)} = \lambda e_{kk} \delta_{ij} + 2\mu e_{ij}, \tag{166}$$

$$\mu_{ij} = -8\mu l^2 \kappa_{ij}. \tag{167}$$

Then, by using Eq. (148), we obtain

$$\sigma_{[ji]} = 2\mu l^2 \varepsilon_{ijk} \nabla^2 \omega_k, \tag{168}$$

for the skew-symmetric part of the force-stress tensor. Therefore, the total force-stress tensor becomes

$$\sigma_{ji} = \lambda e_{kk} \delta_{ij} + 2\mu e_{ij} + 2\mu l^2 \varepsilon_{ijk} \nabla^2 \omega_k. \tag{169}$$

Interestingly, for the equilibrium equation in terms of the displacement, we obtain exactly the same equation as Eq. (62) in MTK theory, that is

$$\left[ \lambda + \mu\left(1 + l^2\right)\nabla^2 \right] u_{k,ki} + \mu(1 - l^2 \nabla^2)\nabla^2 u_i + F_i = 0. \tag{170}$$

However, it should be emphasized that the determinate couple stress theory for isotropic material cannot be taken as a special case of indeterminate MTK theory obtained by letting

$$\eta' = -\eta, \tag{171}$$

and ignoring the indeterminacy term (i.e., $Q = 0$). This is not mathematically valid, because the indeterminacy $Q$ cannot simply be ignored in such an arbitrary manner, and the special case Eq. (171) is excluded specifically in condition Eq. (52) for the MTK couple stress theory. Furthermore, this apparent peculiar coincidence is only for isotropic material. As we see, there is no such analogy for general anisotropic and bi-anisotropic cases. This is just a coincidence for the linear isotropic case, where general equations in both theories have some similarities. It should be realized that the determinate couple stress theory is not simply about fixing the constitutive relations for linear isotropic indeterminate couple stress theory of MTK. This theory is the consistent couple stress theory in continuum mechanics independent of material behavior. This has been achieved by discovering the skew-symmetric character of the couple-stress tensor for continua regardless of isotropic or anisotropic, elastic or inelastic, linear or non-linear properties of material without any additional assumption. This is what Mindlin, Tiersten and Koiter did not recognize in their quest for a consistent couple stress continuum mechanics. The present consistent theory not only resolves the difficulties with former theories, but also provides a powerful tool to develop new theories for different coupled multi-physics problems, such as piezoelectricity [29] and thermoelasticity [30]. This theory has also been introduced into fluid mechanics to model size-dependency and perhaps to contribute to the understanding of turbulence, which affects a cascade of length scales [31].

As mentioned, for isotropic elastic materials, this development is quite remarkable, because only one length scale parameter is needed in addition to the conventional Lamé constants. The results from MTK indeterminate theory for plane isotropic problems [15-22] do remain useful within the determinate couple stress elasticity. However, in MTK theory, recall that there is an extra material constant, along with indeterminacy in the spherical part of the couple-stress tensor.

Interestingly, in recent work based on a discrete site-bond model, Morrison et al. [32] show that a pure twist does not associate with any couple stresses. This confirms the predicted results from determinate couple stress theory that for torsion of an isotropic elastic cylinder, no couple-stresses appear [13].

## 8  Discussion

Table 1 shows the number of elastic coefficients for general linear bi-anisotropic elastic materials in the different generalized size-dependent couple-stress continuum theories. The determinate couple stress theory requires only 27 and 45 elastic constants for the anisotropic and bi-anisotropic cases, respectively. These are less than those in other indeterminate theories, as seen from the table. As mentioned, this character is even more profound in coupled multi-physics problems, which involve new material properties. This can be seen for size-dependent couple stress piezoelectricity and thermoelasticity [29,30]. Thus, the determinate couple stress theory not only is the only fully self-consistent theory, but also is the most parsimonious of the non-classical couple stress theories in terms of independent elastic coefficients.

Interestingly, in the mid-twentieth century, Nobel Laureate C.V. Raman and colleagues [33-35] predicted 45 elastic coefficients for a general anisotropic material and four coefficients for cubic materials, based upon their deep understanding of the spectrum of single crystals.  Although their proposed theoretical development suffers from various difficulties, it is a thought-provoking coincidence that the determinate couple stress theory also predicts exactly the same number of elastic coefficients for the general anisotropic material, as well as for all other symmetry classes including the cubic and isotropic cases.  Table 2 summarizes the number of elastic constants for linear elastic isotropic materials within the different generalized size-dependent couple-stress continuum theories.

**Table 1** Number of elastic coefficients in different linear bi-anisotropic couple stress theories

| Theory | Number of independent measures of deformation | $A_{ijkl}$ | $B_{ijkl}$ | $C_{ijkl}$ | Total |
|---|---|---|---|---|---|
| Indeterminate couple stress theory (MTK) | 14 | 21 | 36 | 48 | 105 |
| Modified couple stress theory (YCLT) | 11 | 21 | 15 | 30 | 66 |
| Indeterminate micropolar theory | 17 | 45 | 36 | 72 | 153 |
| Consistent couple stress theory | 9 | 21 | 6 | 18 | 45 |

**Table 2** Number of elastic coefficients in different linear isotropic couple stress theories

| Theory | Number of independent measures of deformation | $A_{ijkl}$ | $B_{ijkl}$ | $C_{ijkl}$ | Total |
|---|---|---|---|---|---|
| Indeterminate couple stress theory (MTK) | 14 | 2 | 2 | 0 | 4 |
| Modified couple stress theory (YCLT) | 11 | 2 | 1 | 0 | 3 |
| Indeterminate micropolar theory | 17 | 3 | 3 | 0 | 6 |
| Consistent couple stress theory | 9 | 2 | 1 | 0 | 3 |

## 9 Conclusions

In this paper, we have clarified the concept of continuum and demonstrated that its kinematics must be defined only by the continuous displacement field $u_i$, which represents the drift motion of individual particles, regardless of their random motion. This means that the tensors measuring the deformation of the continuum are derived exclusively from this vector displacement field $u_i$ without recourse to any additional independent continuous quantity, such as microrotation $\phi_i$. Therefore, the rotation field $\omega_i$ in a continuum is derived from the displacement field $u_i$, where

$$\omega_i = \frac{1}{2}\varepsilon_{ijk}u_{k,j}. \tag{118}$$

This, in turn, establishes that in a consistent couple stress theory for continuum mechanics:

1. The displacement field $u_i$ and its corresponding rotation field $\omega_i$ are the only degrees of freedom at each point. This means that the rigid body motion of each infinitesimal element of matter at any point of the continuum is described by six degrees of freedom (i.e., three translational $u_i$ and three rotational $\omega_i$).

2. The body couple $C_i$ is not distinguishable from the body force $F_i$. The body couple $C_i$ is merely the result of an equivalent body force and force traction.

3. The concept of generalized force corresponding to independent generalized degrees of freedom requires that the normal component of the surface moment traction vector $m_i^{(n)}$ vanishes everywhere in the continuum.

4. The couple-stress tensor is skew-symmetric and the skew-symmetric part of $\omega_{i,j}$, the mean curvature tensor, is the consistent measure of deformation.

Mindlin and Tiersten realized the importance of considering the rotation $\omega_i = \frac{1}{2}\varepsilon_{ijk}u_{k,j}$ (later called constrained rotation or macrorotation) as the sole rotation field in a continuum. However, the indeterminacy of the couple strees-tensor in their theory was so troublesome that Mindlin developed several other theories, including micropolar theories. Furthermore, in Mindlin and Tiersten [4], body couples appear in the constitutive relations for the force-stress tensor. All of this, of course, severely hampers the applicability of these theories and brings into doubt the existence of such stress components. Yang et al. [6] tried to develop a simpler theory by enforcing a symmetric character on the couple-stress tensor. However, in reality the couple-stress tensor is not symmetric. As a result, their development not only inherits all inconsistencies from indeterminate MTK theory, but also suffers from the inclusion of an artificial equilibrium of moment of couples to the set of fundamental equations. This equilibrium equation has no physical standing and has been created solely to make the couple stress tensor symmetric.

The inability of Mindlin, Tiersten and Koiter to remove the indeterminacy in the original couple stress theory caused many researchers to return to the Cosserat theory, and to consider the independent microrotation $\phi_i$ as a reality. Micropolar theories apparently cure the indeterminacy problem and other issues [9]. However, as we discussed, the artificial microrotation $\phi_i$ as an independent variable is not compatible with the idea of continuous media. Surprisingly, we realize that the pseudo character of microrotation $\phi_i$ still makes the couple stress tensor indeterminate. Moreover, micropolar theories, and their more general form as the microstretch and micromorphic theories, require more material constants in constitutive relations. We now realize that the difficulties of indeterminate MTK theory have been compounded also by increasing the number of material properties.

The skew-symmetric character of the couple-stresses was never recognized by the Cosserats, Mindlin, Koiter, Nowacki, Truesdell and the others, yet this explains why the previous theories suffer from various inconsistencies. In particular, the skew-symmetric character of the couple-stress tensor is the key to resolve all difficulties in MTK theory. We must realize, however, that this does not imply that the developments of Mindlin and Tiersten [4] and Koiter [5] have been in vain. Instead, the new determinate couple stress theory should be considered the natural evolution of the works of Mindlin and Tiersten [4] and Koiter [5], and of the quest for such a theory by Voigt [2] and the Cosserats [3].